\documentclass{cargese}\usepackage{graphicx}
\let\footnote\savefootnote
\let\footnotetext\savefootnotetext 
\let\footnoterule\savefootnoterule 
\normallatexbib

\def\su{SU(1,1)}
\def\ads{{{ADS}$_3$}}

\def\np#1#2#3{Nucl. Phys. {\bf{B#1}} (#2) #3}

\def\plo#1#2#3{Phys. Lett. {\bf{#1B}} (#2) #3}
\begin{document}


\articletitle[On \ads\ string theory]{On \ads\ string
theory\footnote{CPTH-PC-735.0999/September 1999}}


\author{P. Marios Petropoulos}

\affil{Centre de Physique Th\'eorique \\ Ecole Polytechnique \\F-91128
Palaiseau CEDEX}         

\email{marios@cpht.polytechnique.fr}

\begin{abstract}
String theory on curved backgrounds has received much attention
on account of both
its own interest, and of its relation with gauge theories.
Despite the
progress made in various directions, several quite elementary questions
remain unanswered, in particular in the very simple case of
three-dimensional anti-de Sitter space.
I briefly review these problems.
\end{abstract}

String theory is the most appropriate setting for studying
quantum-gravity phenomena. 
In the absence of a truly non-perturbative approach, the usual method
consists in analysing the string propagating on non-trivial backgrounds,
that satisfy the requirements of conformal invariance of the
corresponding sigma model.

Three-dimensional anti-de Sitter
space was recognized long ago as a
case of interest 
\cite{bfow}--\cite{gaw}. It is a
maximally symmetric solution of Einstein's equations with negative
cosmological constant, and time is embedded non-trivially in the curved
geometry. Alternatively, it corresponds to the Freedman--Gibbons
electrovac
solution of gauged supergravity, which can be shown to leave space-time
supersymmetry unbroken \cite{abs}. Other peculiar features of \ads\  
are the absence of
asymptotically flat regions, the presence of boundaries, as well as a rich
causal structure, which makes it possible to obtain three-dimensional
black holes after modding out some discrete symmetry \cite{btz}. 

As far as string theory is concerned, \ads\
is an {\it exact} background, provided
an NS--NS two-form is switched on. The underlying two-dimensional
theory is a
Wess--Zumino--Witten model on the $\su $ group manifold.
Three-dimensional anti-de
Sitter space is 
{\it the only} exact background -- with a single time direction --
where string propagation leads to a WZW model. 	This
property does not hold for more general  ADS$_n$.

Despite many efforts, driven in particular by the recent activity on
ADS/CFT correspondence, 
and the apparent simplicity of the model at hand,
several important and elementary issues
are still beyond our understanding. 

The analysis of string theory on \ads\ plus torsion background
can be performed in two steps. We first study the sigma model
whose target space has the above geometry.
Then, the latter has to be coupled to
two-dimensional gravity. At the level of the Hilbert space, this
amounts to the decoupling of a certain subspace, which becomes
unphysical. Several questions must then be answered, among which: Is the
physical spectrum free of negative-norm states? Is the operator algebra
well-behaved, and what are the vertex operators and their interpretation?
What is the one-loop modular-invariant partition function? 

Very little is known about WZW models on non-compact groups, at a
sufficiently rigorous and general level. Most of our knowledge is based
on a formal extension of the compact case to some specific situations,
and in the framework of current-algebra techniques.
Target-space boundary conditions, in particular, are treated somehow carelessly,
although we know how important they are for selecting various  
representations when studying quantum mechanics on {\ads}~\cite{bs}.

We usually assume that the $SO(2,2)\cong \su_{\rm L} \times \su_{\rm R}$
symmetry of the above model is realized in terms of an affine Lie  
algebra, the level of which is not quantized because $\pi_3(\su)=0$.
The world-sheet energy--momentum tensor is
given by the affine Sugawara construction, and its modes satisfy the
Virasoro algebra
with central charge
$c={3k\over k+2}$. Choosing $k<-2$ leads to one
time and two space directions, while the central charge is positive.

Finally, the Hilbert space is formally constructed as in the compact
case: it is a direct sum of products of representations of the left
and right current algebras.
Highest-weight representations of
the $\su$ current algebra are labelled by the spin $j$ of the primary
fields (states of level zero), which form a representation of the
global algebra. Higher-level states are obtained by acting with negative-frequency 
current oscillators on the above primary states, 
which are annihilated by all positive-frequency modes. 

Irreducible representations of the global algebra are of two
kinds: discrete  or continuous. The discrete ones have highest
or lowest weight, whereas the continuous ones do not. They are all 
infinite-dimensional, at least when unitarity is demanded. This is due to
the non-compact nature of $\su$, which also implies that the metric on
the group manifold is not positive-definite. 

Representations of the $\su$ current algebra have always infinite
towers of negative-norm states. Therefore the corresponding WZW model
cannot be unitary. Nevertheless, since we are interested in the string
theory, the relevant question is whether the physical spectrum is free of
negative-norm states. The latter is defined as the subspace of Virasoro
primaries, satisfying the mass-shell condition, at criticality, namely
when $c_{\rm int} + {3k\over k+2}=26$ ($c_{\rm int}>0$, is the
contribution of some internal CFT, and we are considering here the
bosonic case). Within the present framework, the following conclusions can
be drawn: (\romannumeral1) representations based on continuous series
contain only positive-norm states, but they all describe tachyons; 
(\romannumeral2) representations based on discrete series contain
tachyons, massless and massive excitations, but for sufficiently large 
$\vert j \vert $ ($j$ is negative so as to ensure unitarity of
level-zero states), negative-norm states appear, which remain in the
physical spectrum. 
 
One possible solution to this problem is to cut the spin, hoping that all
negative-norm states are thus eliminated \cite{pmp,moh}. A refined
analysis
shows that a
reasonable constraint is 
$$
{k\over 2}\leq j<0\; .
$$ 
Straightforward norm computations up to level two indicate that unitarity could be
recovered \cite{pmp}, and there are even indications that a no-ghost
theorem could exist \cite{hegp}.
However, an important issue is whether the above condition is consistent with other
features of the string. The answer to this question is not known. Clearly
such a constraint on the allowed values of the spin introduces a cut-off
on the mass of string excitations. This is in obvious disagreement
with
the usual infinite tower of masssive states, and, to some extent, with
what we have learned so far from the ADS/CFT conjecture. 

Another relevant and related question we must face concerns the one-loop partition
function. Several problems appear there. In order to avoid the divergence
due to the infinite degeneracy level by level, we must switch on an
external source coupled to the Cartan generator. This allows for a proper
-- though not  physically clear -- definition of the characters of the
discrete series \cite{kw}. It does not apply, however, succesfully to
continuous
series, whose contribution to the partition function therefore remains
undetermined. Ignoring that contribution, and concentrating on the
discrete series, is not very conclusive either. On the one hand, one
can
seriously argue that, within the above unitarity constraint, modular
invariance cannot be reached \cite{tmr}. On the other hand, even ignoring
the unitarity bound, modular-invariant combinations of the characters are
known when $k>-2$, but it is technically difficult to go beyond, since
the characters themselves have not been computed, except for the case of 
irreducible representations. 

As a conclusion, one can summarize the situation as follows: in the framework of the
current-algebra approach, unitarity seems to be incompatible with a consistent physical
spectrum, since it introduces a cut-off over the mass. This may be interpreted as a
signal  that,
in the present approach, we are missing other
sectors, which could provide us with positive-norm states, whose spin is
$j<k/2$, and which might originate from a proper treatment of the {\it
target-space boundary conditions}. The latter are hard to implement within the
current-algebra method, but might be successfully explored following a path-integral
approach {\it \`a la}  Gaw\c{e}dzki~\cite{gaw}. The role
of continuous series could also be clarified in this way. 
Although they are compatible with unitarity, the corresponding
excitations seem to be all tachyonic. It is fair to mention here that one could try
alternatively
to avoid
the mass/spin cut-off and the purely tachyonic continuous
representations in various ways, playing essentially with the
current algebra, and/or modifying the affine Sugawara
construction \cite {bars}. However, it is not clear
whether such modifications leave unaltered the interpretation of the
theory as a string propagating over \ads.

The motivations for studying the string on \ads\ turn out to be wider than expected in
the early works. The analysis of the
ADS/CFT
conjecture, in the framework of the
\ads $\times S^3$ background, in particular, has attracted
most attention. Ideally, we would like to compute correlators in both
sides and compare them. In practice, correlators for \ads\ string states
are out of reach, which
makes any rigorous check quite
intricate. Therefore, most of the
work in that direction has been devoted to  trying to express the
space-time as well as the
asymptotic two-dimensional conformal symmetry, in terms of the fields
of the WZW model
whose target space is the bulk \ads\ theory, and to build in
that
way the boundary conformal field theory. It is fair to say, however, that
this approach has not shed any light on the structure of the \ads\
string itself
-- at least regarding the questions raised here;
as long as the \ads\ side is not handled exactly, the
achievements
are limited both on checking the ADS/CFT correspondence and on building
the boundary theory \cite{boundconf}. Of course, there still is the
-- weaker -- alternative
to work with the low-energy supergravity, supplemented with all  
Kaluza--Klein excitations coming from higher dimensions, thus trying to
obtain some feedback for the string on \ads.
For example, there are signs that
all $\su$ representations -- discrete and continuous -- should appear
without bound on the mass. If such a bound were present,
assuming the ADS/CFT correspondence,
it
would be hard to identify states
in the bulk \ads\ supergravity (or string, as a fundamental theory), with
states in the boundary conformal field theory.

\begin{acknowledgments}
It is a pleasure to thank C. Bachas and P. Bain for the stimulating  
collaboration we have been having on this subject. 
I would like to thank also the organizers of the Carg\`ese 99 ASI.
The present note is a strict summary of a more extended version
\cite{tmr}, where 
a complete bibliography can be found.
This work was supported in part by EEC TMR contract 
ERBFMRX-CT96-0090.
\end{acknowledgments}



%

\end{document}